\documentclass[
journal=jctcce, % for undefined journal
manuscript=article]{achemso}

\usepackage[version=3]{mhchem} % Formula subscripts using \ce{}
\usepackage{natbib}
\usepackage{graphicx}
\usepackage{subcaption}
\usepackage{longtable} 
\usepackage{notes2bib}
\usepackage{url}
\usepackage{array}
\usepackage{hyperref}
\usepackage{multirow}
\usepackage{marginnote}
\usepackage{bm}
\usepackage{physics}
\usepackage{braket}

\title{First-principles Studies of Strongly Correlated States in Defect Spin Qubits in Diamond}

\author{He Ma}
\affiliation
{Department of Chemistry, University of Chicago, Chicago, Illinois 60637, United States}

\author{Nan Sheng}
\affiliation
{Department of Chemistry, University of Chicago, Chicago, Illinois 60637, United States}

\author{Marco Govoni}
\affiliation
{Pritzker School of Molecular Engineering, University of Chicago, Chicago, Illinois 60637, United States}
\alsoaffiliation
{Materials Science Division and Center for Molecular Engineering, Argonne National Laboratory, Lemont, Illinois 60439, United States.}
\email{mgovoni@anl.gov}

\author{Giulia Galli}
\affiliation
{Pritzker School of Molecular Engineering, University of Chicago, Chicago, Illinois 60637, United States}
\alsoaffiliation
{Department of Chemistry, University of Chicago, Chicago, Illinois 60637, United States}
\alsoaffiliation
{Materials Science Division and Center for Molecular Engineering, Argonne National Laboratory, Lemont, Illinois 60439, United States.}
\email{gagalli@uchicago.edu}

\begin{document}

\begin{abstract}

Using a recently developed quantum embedding theory, we present first principles calculations of strongly correlated states of spin defects in diamond. Within this theory, effective Hamiltonians are constructed, which can be solved by classical and quantum computers; the latter promise a much more favorable scaling as a function of system size than the former. In particular, we report a study of the neutral group-IV vacancy complexes in diamond, and we discuss their strongly-correlated spin-singlet and spin-triplet excited states. Our results provide valuable predictions for experiments aimed at optical manipulation of these defects for quantum information technology applications.

\end{abstract}

\section{Introduction}

Electron spins in molecular and condensed systems are important resources for the storage and process of quantum information \cite{Weber2010}. In the past decades, several spin-defects in wide band gap semiconductors and insulators have been widely studied, in particular in diamond \cite{Doherty2013}, silicon carbide \cite{Weber2011,Christle2015}, and aluminum nitride \cite{Seo2016,Seo2017}. The prototype example of spin-defects is the negatively-charged nitrogen-vacancy center (NV) center in diamond \cite{Davies1976,Rogers2008,Doherty2011,Maze2011,Choi2012,Goldman2015}. The NV center exhibits spin-triplet ground state with long spin coherence time even at room temperature \cite{Balasubramanian2009}. Different spin states of the electron spin can be used to encode quantum information, and transitions between spin states can be driven by microwave fields. To date, spin-defects have found many applications both in fundamental science and cutting-edge quantum technologies. For instance, spin-defects have been used to demonstrate fundamental principles of quantum mechanics such as the Berry phase \cite{Yale2016} and Bell inequality \cite{Hensen2015}. Spin-defects are also extensively used as quantum sensors due to their sensitivity to external electric, magnetic and temperature fields \cite{Hsieh2019,Fukami2019}. Furthermore, the spin states of defects can be coupled with various optical \cite{Morse2017} and mechanical \cite{Whiteley2019} degrees of freedom, making them important components in hybrid quantum architectures for quantum communication and quantum computation.

First-principles simulations based on density functional theory (DFT) have been playing an important role in the identification and characterization of spin-defects \cite{Seo2016,Seo2017}. For instance, ground state DFT calculations can predict the formation energies of defects, thus enabling, e.g. the identification of the atomistic structure and charge states of unknown defects \cite{Ivady2018}. Using ground state DFT wavefunctions, several spin properties can be computed that are critical for the prediction of qubit state splitting and coherence time, such as the zero-field splitting and the hyperfine coupling \cite{Ghosh2019,Ma2020pyzfs}. However accurate predictions of excited states are challenging, when using DFT, especially in the case of strongly correlated states which may not be approximated by a single Slater determinant of spin-electron orbitals. Multi-reference electronic states have been an important subject of research in  quantum chemistry for decades \cite{Helgaker}. Unfortunately, most \textit{ab initio} multireference methods are computationally very demanding, preventing their straightforward application to spin-defects in solids, whose description requires periodic supercells containing hundreds of atoms.

In the past decades, quantum embedding theories emerged as promising approaches to apply a high-level theory (such as multireference methods) to the description of strongly correlated active regions of a solid or molecule, where the environment is treated with a lower level of theory. Different quantum embedding schemes have been proposed\cite{Sun2016}, using, e.g. the electron density \cite{Huang2006,Huang2011,Goodpaster2014,Jacob2014,Genova2014,Wen2019}, density matrices \cite{Knizia2012,Wouters2016,Pham2019} or based on Green’s function approaches\cite{Nguyen2016,Dvorak2019,Zhu2019,Aryasetiawan2004,Aryasetiawan2009,Miyake2009,Imada2010,Hirayama2013,Hirayama2017}. Spin-defects in semiconductors can be viewed as atom-like systems embedded in bulk crystals, and the states used to encode quantum information are usually localized around the defects. Therefore, spin-defects are promising systems for the application of quantum embedding theories. For instance, Bochstedte and coworkers investigated strongly correlated excited states of NV in diamond and divacancies in silicon carbide using the constrained random phase approximation (cRPA) \cite{Bockstedte2018}. In the cRPA approach \cite{Aryasetiawan2004,Miyake2009,Hirayama2017}, the low-energy excited states of the active site are obtained by solving an effective Hamiltonian that is constructed from effective electron-electron interactions. The cRPA approach is based on the random phase approximation (RPA), which neglects exchange-correlation effects in the calculation of dielectric screening. Recently, we developed a quantum embedding theory \cite{Ma2020} similar to cRPA, albeit going beyond the RPA description of dielectric screening by including exchange-correlation effects evaluated using a finite-field algorithm \cite{Ma2018,Nguyen2019}. In addition, the quantum embedding theory of Ref.\cite{Ma2020} has the important advantage that no explicit summation over empty electronic orbitals is necessary \cite{Wilson2008,Nguyen2012,Pham2013,Govoni2015}, making it scalable to systems with hundreds of atoms. We demonstrated the efficiency and accuracy of such a computational approach for spin-defects in diamond and silicon carbide, and carried out calculations on both classical and quantum computers.

\begin{figure}[H]
  \centering
  \includegraphics[width=5in]{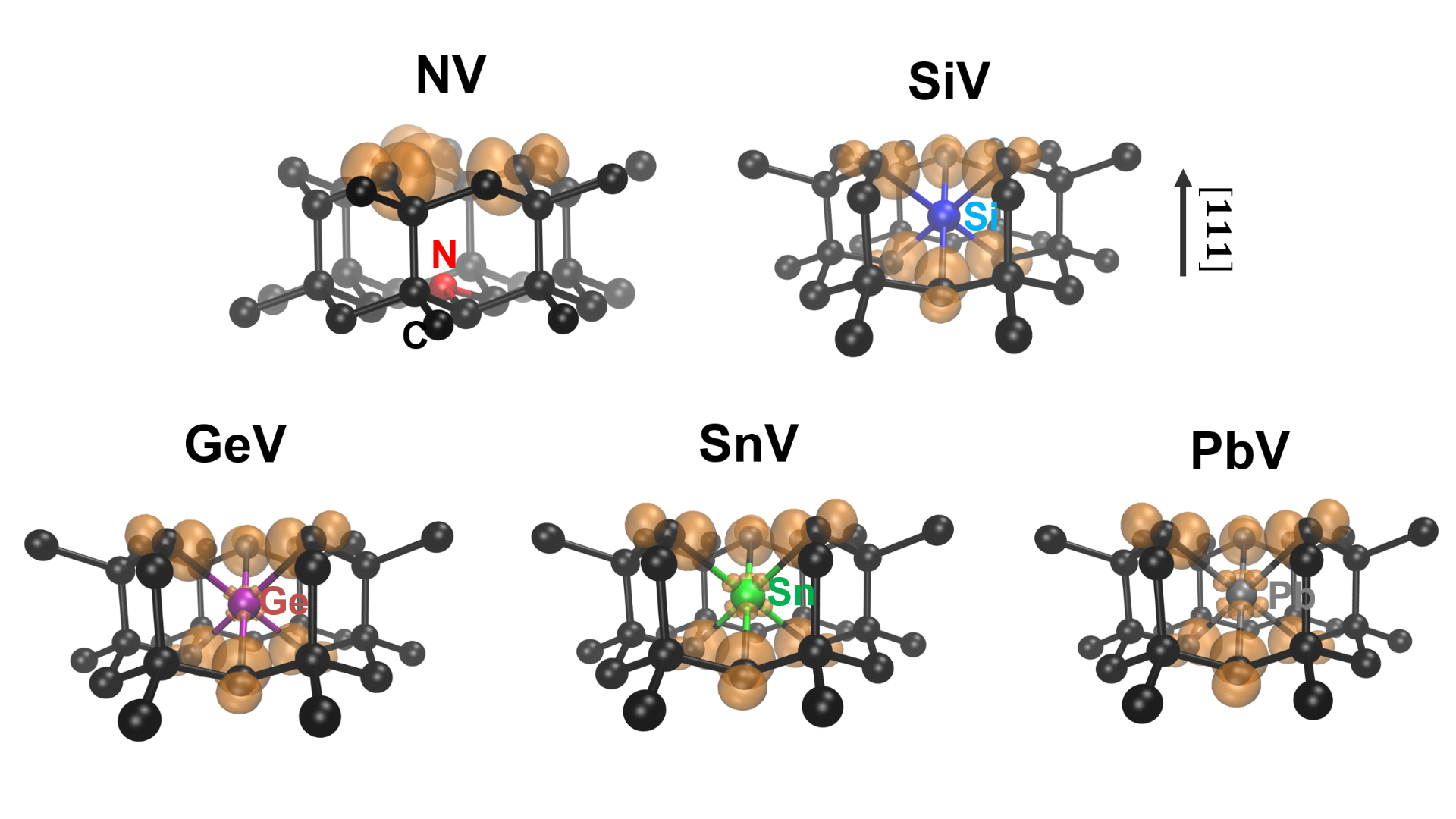}
  \caption{Structures and spin polarization densities of spin-defects in diamond, including the negatively-charged nitrogen-vacancy (NV) center, and the neutral group-IV vacancy complexes XV (with X=Si, Ge, Sn, and Pb).}
  \label{structures}
\end{figure}

In this work, we apply the quantum embedding theory of Ref.\cite{Ma2020} to several defects in diamond (Fig. \ref{structures}). In particular, we consider the group-IV vacancy complexes in diamond, i.e. XV where X=Si, Ge, Sn, Pb, in addition to the NV center. These vacancy complexes have attracted substantial interests recently due to their excellent optical properties \cite{Haenens2011,Gali2013,Thiering2018,Green2019,Thiering2019,Zhang2020}. We performed quantum embedding calculations based on DFT results obtained with different exchange-correlation functionals \cite{Perdew1996,Skone2014} and demonstrated the importance of using hybrid functionals to obtain accurate results.  While the NV and SiV centers were discussed in part in ref. \cite{Ma2020}, here we report the first to-date simulation of the strongly-correlated excited states of the neutral GeV, SnV and PbV defects in both the spin singlet and spin triplet manifold, which are both required to predict their operation as optically-addressable qubits.

\pagebreak

\section{Methods}

\subsection{Quantum embedding theory}

For a system of interacting electrons, the non-relativisitic Hamiltonian is given by
\begin{equation} \label{Hfull}
    H = \sum_{ij} t_{ij} a_{i}^{\dagger} a_{j} + \frac{1}{2} \sum_{ijkl} v_{ijkl} a_{i}^{\dagger} a_{j}^{\dagger} a_{l} a_{k}
\end{equation}
where $a^{\dagger}$ and $a$ are creation and annihilation operators acting on single-electron orbitals $i, j, k, l$; the one-electron term $t$ includes the kinetic energy and the electron-nuclei interaction; the two-electron term $v$ represents the \textit{bare} Coulomb interaction between electrons. The exact solution of $H$ is generally limited to small systems due to the high computational cost. 

For systems where important electronic excitations are restricted to an active space (A) such as frontier orbitals of molecules or energy levels near the Fermi level of solids, it is desirable to construct an effective Hamiltonian that operates only on the active space 
\begin{equation} \label{Heff}
    H^{\mathrm{eff}} = \sum_{ij}^{\mathrm{A}} t^{\mathrm{eff}}_{ij} a_{i}^{\dagger} a_{j} + \frac{1}{2} \sum_{ijkl}^{\mathrm{A}} v^{\mathrm{eff}}_{ijkl} a_{i}^{\dagger} a_{j}^{\dagger} a_{l} a_{k}.
\end{equation}
and the physical processes outside the active space are included through a renormalization of $t$ and $v$. The renormalized effective Hamiltonian parameters $t^{\mathrm{eff}}$ and $v^{\mathrm{eff}}$ should properly incorporate dielectric screening and exchange-correlation effects outside the active space. In the cRPA approach, the two-body term in the effective Hamiltonian $v^{\text{eff}}$ is computed as a partially screened Coulomb interaction
\begin{equation}
    v^{\mathrm{eff}} = v + v \chi_{\mathrm{rpa}}^\mathrm{E} v
\end{equation}
where $\chi_{\mathrm{rpa}}^\mathrm{E} = \chi_0^\text{E} + \chi_0^\text{E} v \chi_{\mathrm{rpa}}^\mathrm{E}$ is the reducible polarizability of the environment within the RPA; $\chi_0^\text{E} = \chi_0 - \chi_0^\text{A}$ is the irreducible density response function for the environment (E), with $\chi_0^\text{A}$ being the projection of $\chi_0$ inside the active space.

The cRPA approach neglects the exchange-correlation effect in the calculation of the dielectric screening. 
%$W^\text{E}_{\text{rpa}}$. 
In Ref.  \cite{Ma2020}, we proposed an expression for $v^{\mathrm{eff}}$ that properly accounts for exchange and correlation interactions in the environment
\begin{equation} \label{Veff}
    v^{\mathrm{eff}} = v + f \chi^\mathrm{E} f
\end{equation}
where the reducible density response function $\chi^\mathrm{E}$ of the environment is evaluated beyond the RPA as $\chi^\mathrm{E} = \chi_0^E + \chi_0^\mathrm{E} f \chi^E$, with $f = v + f_{\mathrm{xc}}$ being the Hartree-exchange-correlation kernel. The exchange-correlation kernel $f_{\mathrm{xc}}$, defined as the derivative of the exchange-correlation potential with respect to the electron density, is evaluated with a finite-field algorithm described in Ref. \cite{Ma2018,Nguyen2019}. By representing $\chi^\mathrm{E}$ and $f$ on a compact basis obtained from a low-rank decomposition of the dielectric matrix \cite{Wilson2008,Govoni2015}, one can avoid the evaluation and summation over virtual electronic orbitals. Finally, the one-body term $t^{\mathrm{eff}}$ can be computed by properly subtracting from the Kohn-Sham Hamiltonian a term that accounts for Hartree and exchange-correlation effects in the active space \cite{Ma2020}.

\subsection{Computational setup}

\begin{figure}[H]
  \centering
  \includegraphics[width=6in]{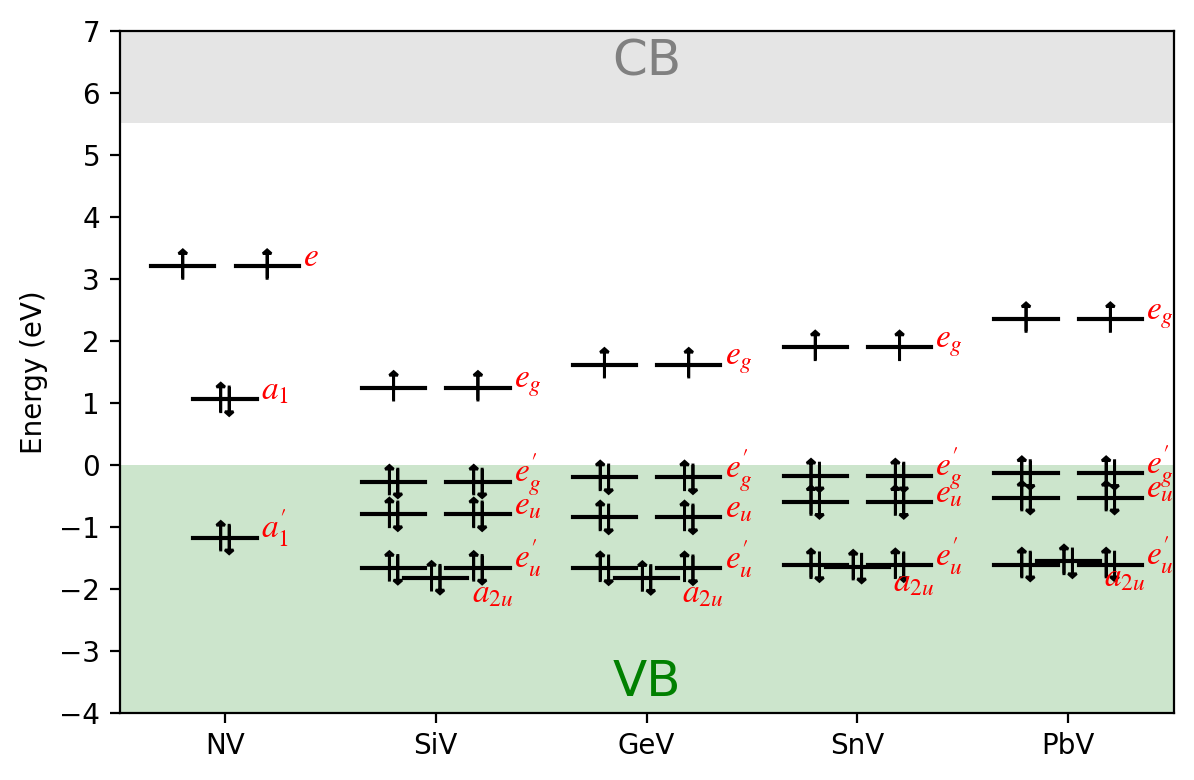}
  \caption{Mean-field electronic structure of spin-defects in diamond obtained with spin-restricted DFT calculations using the dielectric-dependent hybrid functional (DDH)\cite{Skone2014}. VB (CB) denotes the valence (conduction) band. The symmetry of important defect orbitals is indicated following group theory notation.}
  \label{levels}
\end{figure}

We first carried out spin-unrestricted DFT calculations  to obtain the ground state geometries of defects in their host materials. Using ground state geometries, we then performed spin-restricted DFT calculations \cite{Skone2014} to obtain their electronic structure (Fig. \ref{levels}) at the mean-field level, which serves as the starting point for the construction of the effective Hamiltonian described in the previous section. The spin restriction ensures that both spin channels are treated on equal footing and the eigenstates of the resulting effective Hamiltonian are eigenstates of $S^2$. Once mean-field DFT single particle eigenvalues and wavefunction are obtained, an effective Hamiltonian was constructed using the quantum embedding theory described in Sec. 2.1. The active space is defined by a set of selected Kohn-Sham orbitals, that are chosen to include relevant defect levels in the band gap of the host material, as well as resonance orbitals and orbitals close in energy to band edges. The choice of the  active space was tested to yield converged excitation energies (see Supplementary Information (SI)). Full configuration-interaction (FCI) calculations \cite{Knowles1984} were performed for the effective Hamiltonian to compute low-energy eigenstates and vertical excitation energies.

We performed DFT calculations using the Quantum Espresso code \cite{Giannozzi2009}. We used a plane-wave basis set with a kinetic energy cutoff of 50 Ry. Norm-conserving pseudopotentials from the SG15 library \cite{Schlipf2015} are used to represent electron-ion interactions; these pseudopotentials include scalar relativistic effects. In the calculations of PbV including spin-orbit coupling, we used the fully relativistic version of SG15 pseudopotentials \cite{Scherpelz2016}. Defects are modeled with 215-atom supercells of diamond with the $\Gamma$-point sampling of the Brilliouin zone. Quantum embedding theory calculations are performed from two different DFT starting points, obtained respectively  with PBE \cite{Perdew1996} and a dielectric-dependent hybrid functional (DDH) \cite{Skone2014}, using geometries optimized with the PBE functional. For selected cases we also tested the HSE06 functional \cite{Heyd2003,Krukau2006}, which was found to yield  results similar to those of the DDH functional (see SI). Quantum embedding calculations were carried out with the WEST code \cite{Govoni2015}. Density response functions were evaluated using a basis set including the first 512 eigenvectors of $\chi_0$. In calculations beyond the RPA, the exchange-correlation kernel $f_{\text{xc}}$ was computed with a finite-field algorithm using the WEST code coupled to the Qbox code \cite{Gygi2008}. FCI calculations were carried out using the PySCF code \cite{Sun2017}.

\pagebreak

\section{Results}

In Table \ref{excitation_energies} we summarize several vertical excitation energies of spin-defects obtained from FCI calculations with the Hamiltonian defined in Sec. 2.1. Overall, the excitation energies obtained using the DFT@DDH energies and wavefunctions are significantly larger than those obtained at the DFT@PBE level of theory, and DDH results are in better agreement with available reference values. Our findings highlight the importance of using DFT@DDH as a starting point for embedding calculations.

\begin{longtable}[H]{lllllll}
% \resizebox{\textwidth}{12mm}
\caption{Vertical excitation energies (eV) of spin-defects including the negatively charged nitrogen-vacancy (NV) and neutral silicon-vacancy (SiV), germanium-vacancy (GeV), tin-vacancy (SnV), and lead-vacancy (PbV) center in diamond. Calculations are performed using PBE and DDH functionals to obtain mean-field energy levels, and  dielectric screening is evaluated within and beyond the random phase approximation (RPA). Reference vertical excitation energies are computed from experimental zero-phonon lines (ZPL) when Stokes energies are available. Reference experimental values for ZPLs are shown in brackets in the Ref column.}
% \begin{tabular}{lllllll}
\\
\hline &  & \multicolumn{2}{l}{PBE} & \multicolumn{2}{l}{DDH} &           Ref \\
    &                                        &     RPA & Beyond-RPA &     RPA & Beyond-RPA \\
System & Excitation &         &            &         &            &               \\
\hline
NV & ${}^3E_{} \leftrightarrow {}^3A_{2}$ &  1.395 &      1.458 &  1.921 &      2.001 &  2.180 \cite{Davies1976} (1.945 \cite{Davies1976}) \\
       & ${}^1A_{1} \leftrightarrow {}^3A_{2}$ &  1.211 &      1.437 &  1.376 &      1.759 &                \\
       & ${}^1E_{} \leftrightarrow {}^3A_{2}$ &  0.396 &      0.444 &  0.476 &      0.561 &                \\
       & ${}^1A_{1} \leftrightarrow {}^1E_{}$ &  0.815 &      0.993 &  0.900 &      1.198 &        (1.190 \cite{Rogers2008}) \\
       & ${}^3E_{} \leftrightarrow {}^1A_{1}$ &  0.184 &      0.020 &  0.545 &      0.243 &  (0.344-0.430 \cite{Goldman2015}) \\
\hline
SiV & ${}^3E_{u} \leftrightarrow {}^3A_{2g}$ &   1.247 &      1.258 &   1.590 &      1.594 &  1.568 \cite{Thiering2019} (1.31 \cite{Green2019}) \\
    & ${}^3A_{1u} \leftrightarrow {}^3A_{2g}$ &   1.386 &      1.416 &   1.741 &      1.792 &               \\
    & ${}^1E_{g} \leftrightarrow {}^3A_{2g}$ &   0.232 &      0.281 &   0.261 &      0.336 &               \\
    & ${}^1A_{1g} \leftrightarrow {}^3A_{2g}$ &   0.404 &      0.478 &   0.466 &      0.583 &               \\
    & ${}^1A_{1u} \leftrightarrow {}^3A_{2g}$ &   1.262 &      1.277 &   1.608 &      1.623 &               \\
    & ${}^3E_{u} \leftrightarrow {}^3A_{2u}$ &  -0.000 &      0.002 &   0.003 &      0.011 &       (0.007 \cite{Green2019}) \\
\hline
GeV & ${}^3E_{u} \leftrightarrow {}^3A_{2g}$ &   1.595 &      1.619 &   2.076 &      2.105 &               \\
    & ${}^3A_{1u} \leftrightarrow {}^3A_{2g}$ &   1.689 &      1.726 &   2.173 &      2.231 &               \\
    & ${}^1E_{g} \leftrightarrow {}^3A_{2g}$ &   0.288 &      0.355 &   0.329 &      0.434 &               \\
    & ${}^1A_{1g} \leftrightarrow {}^3A_{2g}$ &   0.529 &      0.639 &   0.617 &      0.797 &               \\
    & ${}^1A_{1u} \leftrightarrow {}^3A_{2g}$ &   1.595 &      1.621 &   2.076 &      2.110 &               \\
    & ${}^3E_{u} \leftrightarrow {}^3A_{2u}$ &  -0.012 &     -0.011 &  -0.012 &     -0.009 &               \\
\hline
SnV & ${}^3E_{u} \leftrightarrow {}^3A_{2g}$ &   1.579 &      1.599 &   2.069 &      2.091 &               \\
    & ${}^3A_{1u} \leftrightarrow {}^3A_{2g}$ &   1.667 &      1.696 &   2.160 &      2.207 &               \\
    & ${}^1E_{g} \leftrightarrow {}^3A_{2g}$ &   0.302 &      0.368 &   0.341 &      0.444 &               \\
    & ${}^1A_{1g} \leftrightarrow {}^3A_{2g}$ &   0.565 &      0.678 &   0.649 &      0.830 &               \\
    & ${}^1A_{1u} \leftrightarrow {}^3A_{2g}$ &   1.570 &      1.591 &   2.060 &      2.086 &               \\
    & ${}^3E_{u} \leftrightarrow {}^3A_{2u}$ &  -0.017 &     -0.017 &  -0.017 &     -0.014 &               \\
\hline
PbV & ${}^3E_{u} \leftrightarrow {}^3A_{2g}$ &   1.910 &      1.934 &   2.464 &      2.493 &               \\
    & ${}^3A_{1u} \leftrightarrow {}^3A_{2g}$ &   1.980 &      2.008 &   2.533 &      2.574 &               \\
    & ${}^1E_{g} \leftrightarrow {}^3A_{2g}$ &   0.321 &      0.396 &   0.360 &      0.476 &               \\
    & ${}^1A_{1g} \leftrightarrow {}^3A_{2g}$ &   0.615 &      0.750 &   0.697 &      0.910 &               \\
    & ${}^1A_{1u} \leftrightarrow {}^3A_{2g}$ &   1.894 &      1.916 &   2.446 &      2.476 &               \\
    & ${}^3E_{u} \leftrightarrow {}^3A_{2u}$ &  -0.023 &     -0.024 &  -0.025 &     -0.025 &               \\
\hline
% \end{tabular}
\label{excitation_energies}
\end{longtable}

\begin{figure}[H]
  \centering
  \includegraphics[width=4in]{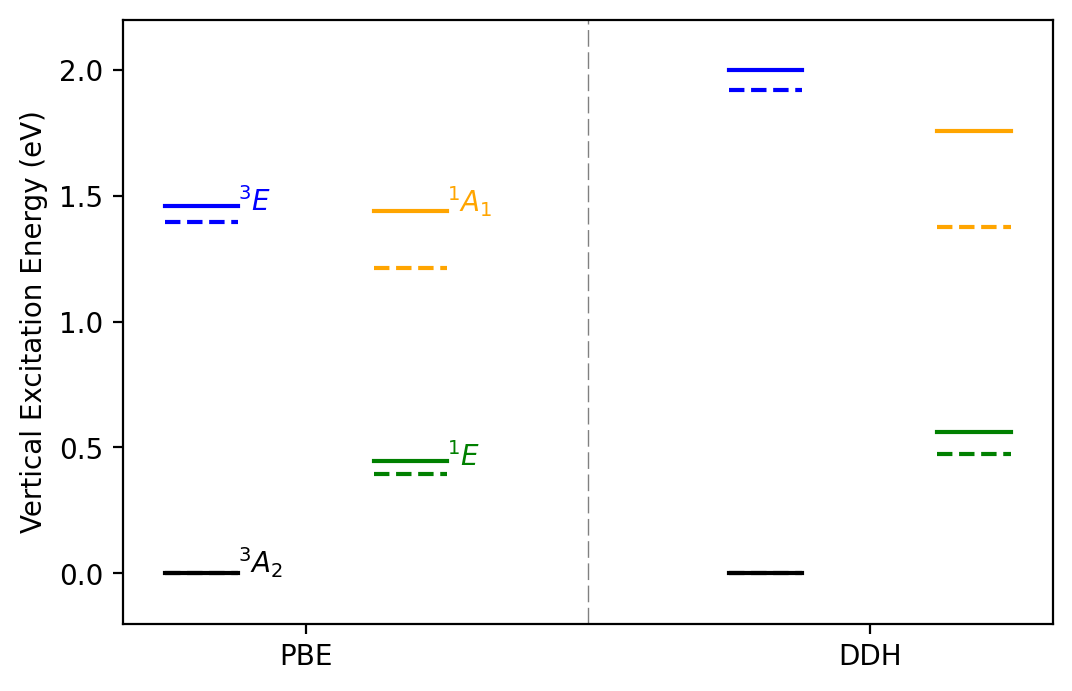}
  \caption{Many-electron energy levels of negatively charged nitrogen-vacancy (NV) center in diamond. Calculations are performed starting from PBE and dielectric-dependent hybrid (DDH) functionals, with dielectric screening evaluated within (dashed lines) and beyond (solid lines) the random phase approximation (RPA).}
  \label{nv}
\end{figure}

The NV in diamond has a spin-triplet ground state of C$_{3v}$ symmetry. Fig. \ref{nv} shows its vertical excitation energies  computed within and beyond the RPA, using the PBE and DDH functionals. In all cases, quantum embedding calculations predict the correct energy level structure of $^3A_2$, $^1E$, $^1A_1$ and $^3E$, with $^1E$ and $^1A_1$ being strongly-correlated states that cannot be directly computed by DFT. Results obtained beyond the RPA using the DDH functional yield the best agreement with experimental values (Table \ref{excitation_energies}).

Group-IV vacancy centers (SiV, GeV, SnV and PbV) in diamond have spin-triplet ground states with D$_{3d}$ symmetry. The spin-flip excitations within $e_g$ single-particle defect levels in the band gap and the excitations from $e_u$ to $e_g$ orbitals yield a rich set of many-electron excited states, many of which are strongly-correlated. Experimentally, it has been shown that the lowest spin-triplet excitations of SiV lead to a $^3A_{2u}$-$^3E_{u}$ manifold \cite{Green2019}. Much less is known about spin singlet excited states. Here we provide the first predictions of the singlet states of GeV, SnV and PbV obtained with first-principles simulations.

\begin{figure}[H]
  \centering
  \includegraphics[width=4in]{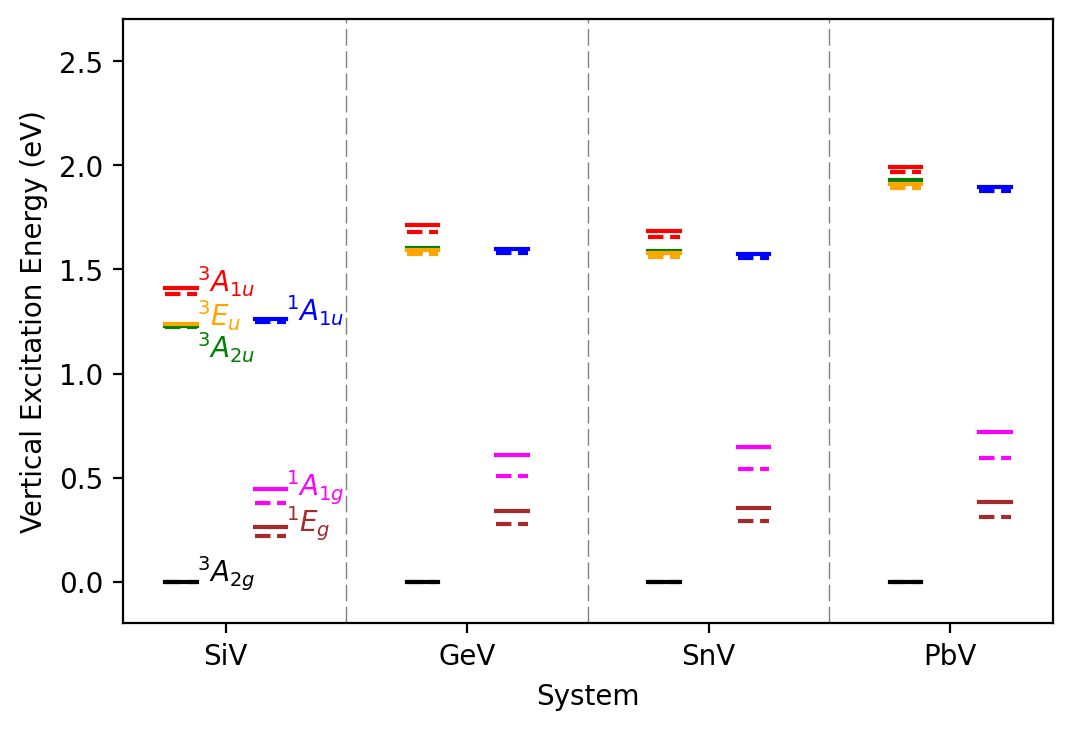}
  \includegraphics[width=4in]{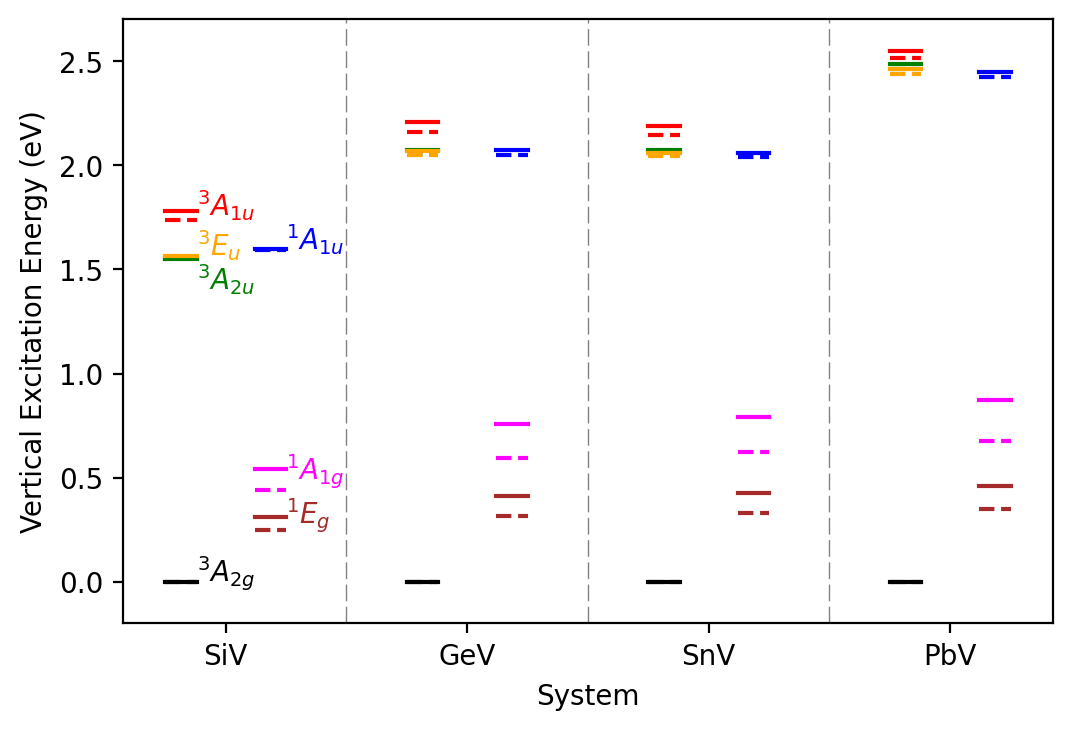}
  \caption{Many-electron energy levels of the neutral silicon-vacancy (SiV), germanium-vacancy (GeV), tin-vacancy (SnV) and lead-vacancy (PbV) center in diamond. Calculations are performed starting from PBE (top) and dielectric-dependent hybrid (DDH) (bottom) functionals, with dielectric screening evaluated within (dashed lines) and beyond (solid lines) the random phase approximation (RPA).}
  \label{siv}
\end{figure}

Fig. \ref{siv} presents the vertical excitation energies of many-electron states of group-IV vacancy centers. First, we note that the excitation energies from $^3A_{2g}$ state to $^3E_{u}$ state increase from SiV to PbV (1.594/2.105/2.091/2.493 eV for SiV/GeV/SnV/PbV), which is consistent with the trend of increasing $e_u$-$e_g$ energy level splitting in their mean-field descriptions (Fig. \ref{levels}). In the spin singlet manifold, the positions of $^1E_{g}$, $^1A_{1g}$ and $^1A_{1u}$ are also increasing in energy from SiV to PbV. These singlet states originate from spin-flip transitions of $e_g$ defect orbitals located in the band gap of diamond, and thus their excitation energies strongly depend on the Coulomb repulsion of electrons in $e_g$ orbitals. The increasing excitation energies indicate an increase in strength of the effective Coulomb interactions, as the element becomes heavier (the bond length between impurity atom and nearest neighbor carbon atom is 1.99/2.03/2.10/2.13 {\rm \AA} for SiV/GeV/SnV/PbV, respectively).

In the case of PbV, we investigated the influence of spin-orbit coupling by performing fully relativistic DFT calculations with noncollinear spin. We found that the effect of spin-orbit coupling (SOC) on the position and splitting of defect levels (see SI) is negligible. For instance, the $e_g$ orbitals of PbV in the band gap of diamond are split by less than 0.02 eV due to the SOC effect. We further carried out projected density of states calculations (see SI), which indicate that defect orbitals are hybrid orbitals with a major component coming from the host carbon atoms instead of the impurity atom. This prominent carbon character of the orbitals is responsible for the small SOC splitting observed in the PbV case. Therefore, we concluded that spin-orbit coupling could be neglected in our quantum embedding calculations.

Comparing results obtained with PBE and DDH functionals, we again found that the DDH functional yields larger excitation energies and is in closer agreement with experiments than those obtained with PBE. Beyond-RPA calculations yield larger singlet excitation energies than those obtained with RPA, similar to our findings for NV. Unlike singlet excitation energies, triplet excitation energies of group-IV vacancy centers are found to be insensitive to the description of dielectric screening and mainly depend on the mean-field starting point.

Experimentally, it has been challenging to realize optical spin polarization for the neutral SiV; however important progress in that direction has been recently  reported by Zhang et al. \cite{Zhang2020}, who performed optically detected magnetic resonance measurements enabled by optical spin polarization via higher-lying excited states. Our results for SiV indicate that the experimental difficulties may arise from the position of the $^1A_{1u}$ state being slightly higher in energy than that of the $^3A_{2u}$-$^3E_{u}$ manifold, thus making the intersystem crossing (ISC) from triplet to singlet manifolds energetically unfavorable \cite{Ma2020}. However, when moving from SiV to PbV in group IV, the the $^1A_{1u}$ state becomes slightly lower in energy than  the $^3A_{2u}$-$^3E_{u}$ manifold, suggesting that the ISC may become energetically more favorable for heavier defects, such as the PbV.

\section{Conclusion}

In summary, we presented a study of strongly-correlated electronic states of several spin-defects in diamond using the quantum embedding theory described in Ref~\cite{Ma2020}. We reported the first prediction of strongly-correlated electronic states of neutral GeV, SnV and PbV defects based on first-principles calculations. In addition, we compared results obtained starting from different functionals and with different approximations in the treatment of the dielectric screening, and we showed the importance of using hybrid functional starting points and beyond-RPA dielectric screening for the construction of effective models of spin-defects. Our results indicate that optical spin polarization may be easier to realize in neutral vacancy complexes with elements heavier than Si, e.g. Pb, due to a more energetically favorable ISC. Finally we note that the quantum embedding results obtained in this work are based on the exact diagonalization of effective Hamiltonians, which can be effectively performed on near-term quantum computers with a relatively small number of qubits, as shown in Ref~\cite{Ma2020}.

\pagebreak

\section*{Conflicts of interest}
There are no conflicts to declare.

\section*{Acknowledgements}
This work was supported by MICCoM, as part of the Computational Materials Sciences Program funded by the U.S. Department of Energy, Office of Science, Basic Energy Sciences, Materials Sciences, and Engineering Division through Argonne National Laboratory, under contract number DE-AC02-06CH11357 and by AFOSR FA9550-19-1-0358. This research used resources of the National Energy Research Scientific Computing Center (NERSC), a DOE Office of Science User Facility supported by the Office of Science of the US Department of Energy under Contract No. DE-AC02-05CH11231 and resources of the University of Chicago Research Computing Center.

\bibliography{ref}

\end{document}